\documentclass[aps,a4paper,nofootinbib]{revtex4}

\usepackage{amsmath,amssymb}
\usepackage{graphicx}
\usepackage{xcolor}
\usepackage{hyperref}
\usepackage{lineno}
\usepackage{xspace}

\newcommand{\dQ}{\ensuremath{{\Delta}Q}\xspace}

\newcommand{\Kt}{\ensuremath{\kappa_2}\xspace}
\newcommand{\Kf}{\ensuremath{\kappa_4}\xspace}
\newcommand{\chiQtwo}{\ensuremath{\chi_{2}^{\dQ}}\xspace}
\newcommand{\chiQfour}{\ensuremath{\chi_{4}^{\dQ}}\xspace}

\newcommand{\pT}{\ensuremath{p_{\rm T}}\xspace}
\newcommand{\kQ}{\ensuremath{k_{\rm Q}}\xspace}
\newcommand{\NXIM}{\ensuremath{N_{+}}\xspace}
\newcommand{\NXIP}{\ensuremath{N_{-}}\xspace}
\newcommand{\mNch}{\ensuremath{\langle N_{\rm ch} \rangle}\xspace}
\newcommand{\pp}{pp\xspace}

\newcommand{\pbpb}{Pb--Pb\xspace}

\newcommand{\sppt}[1]{\ensuremath{\sqrt{s} = #1\,\text{TeV}}\xspace}
\newcommand{\gevc}[1]{\ensuremath{#1\,\text{GeV}/c}\xspace}

\begin{document}

\title{Testing thermal equilibrium in heavy-ion collisions\\
with electric-charge fluctuations}

\author{Peter Christiansen\footnote{peter.christiansen@fysik.lu.se}}

\affiliation{Division of Particle and Nuclear Physics, Department of Physics, Lund University, Lund, 22364, Sweden}

\date{\today}


\begin{abstract}
Relativistic heavy-ion collisions exhibit several phenomena commonly associated
with the formation of a thermalized quark--gluon plasma. However, the observation of similar phenomena in small systems, and their non-thermal microscopic descriptions there, raises questions about this assumption.

Recent continuum-extrapolated lattice-QCD calculations of the second- and fourth-order electric-charge susceptibilities, $\chiQtwo$ and $\chiQfour$, reveal a striking tension with hadron-resonance-gas expectations, particularly for $\chiQfour$. As the ratio $\chiQfour/\chiQtwo$ is experimentally measurable with net-charge cumulants, this opens up completely new ways to test whether thermal QCD applies in central heavy-ion collisions. However, achieving a measurement that can be interpreted with high precision requires further work on both the phenomenological and experimental sides.

 Here, the focus is on an experimental counterpart to recent phenomenological work on the $D$-measure of net-charge fluctuations: how restricted detector acceptance in rapidity and transverse momentum can be controlled experimentally. We demonstrate this for the second-order net-charge cumulant, $\Kt(\dQ)$, by rewriting it in terms of the charge balance function and the charged-particle rapidity density. Both are event-averaged observables for which experimental efficiency corrections and extrapolations into unmeasured kinematic regions can be studied systematically. This provides a complementary experimental route for connecting measurements of $D = 4\Kt(\dQ)/\mNch$ to theory and a possible starting point for extending the same strategy to the fourth-order cumulant.
\end{abstract}

\maketitle

\section{Introduction}
\label{sec:introduction}

The matter produced in central heavy-ion collisions at the RHIC and the LHC is
commonly interpreted as a quark--gluon plasma (QGP)~\cite{Arslandok:2023utm,ALICE:2022wpn,CMS:2024krd} that is in thermal equilibrium in central collisions, see for example Ref.~\cite{Harris:2023tti}.  This interpretation has wide-ranging consequences as it motivates the use of equilibrium lattice-QCD calculations, e.g., for the equation of state, and the use of statistical hadron-resonance-gas (HRG) models to describe hadron production at chemical freeze-out.

However, the observation of QGP-like effects in small systems, see for example Refs.~\cite{Grosse-Oetringhaus:2024bwr,Christiansen:2024bhe}, has provided a very different perspective on the QGP. First, it became clear that hydrodynamics does not require thermalization, see Ref.~\cite{Strickland:2024moq} for a recent review, i.e., a system can develop a hydrodynamic description
without all microscopic degrees of freedom being in thermal
equilibrium. Later it became clear that even hydrodynamics is not applicable in the smallest systems where QGP-like effects are observed~\cite{Kurkela:2018xxd,Ambrus:2022qya}, which warrants new explanations for the observed phenomena there. This has led to different explanations for the observed flow in these systems, among which two prominent examples are kinetic theory~\cite{Kurkela:2018ygx} and Lund string shoving~\cite{Bierlich:2017vhg}. While none of these models are able to explain all the observed phenomena, it is clearly in the interest of the field to confront the underlying ideas and it has recently been understood that this is at least in principle possible in ultra-small systems~\cite{Bierlich:2024lmb}. 

On the large system side, microscopic event generators such as PYTHIA~\cite{Bierlich:2022pfr}, which are highly successful in proton-proton collisions, have been extended to heavy-ion collisions, Angantyr~\cite{Bierlich:2018xfw}, with some success, e.g., giving non-thermal explanations for flow~\cite{Bierlich:2017vhg} and strangeness enhancement~\cite{Bierlich:2014xba}. This raises the point that the current standard of using successful descriptions of observables to validate the underlying physics is at most a ``silver'' standard.

In this paper we propose that fluctuations of conserved electric charge offer
a qualitatively different test.  Conserved-charge susceptibilities are
well-defined equilibrium quantities calculable from the QCD partition
function.  A recent lattice-QCD calculation has, for the first time, obtained
the fourth-order electric-charge susceptibility in the continuum limit over
the temperature range $T=120$--$160$\,MeV, together with a precise calculation
of the second-order susceptibility~\cite{Borsanyi:2026jbj}.  The resulting ratio
$\chiQfour/\chiQtwo$ differs markedly from hadron-resonance-gas (HRG)
expectations in the temperature region relevant for heavy-ion collisions.
The lattice authors explicitly propose a measurement of this ratio at the LHC, where the same observable can be measured via charge-fluctuation cumulant ratios. 

The purpose of the present work is to place this proposal in a broader context:
rather than viewing the measurement only as a test of the treatment of pion
interactions in the HRG, we ask whether it can be developed into a direct
experimental consistency test of the assumption of thermal equilibrium in
central heavy-ion collisions. Net-charge fluctuations appear to be highly suitable for this as these are observables where both experiment and theory have many strengths. 1) There is a long experimental tradition of measuring similar observables via cumulants. 2) Electric charge can be measured more directly than baryon number so it seems that this is an observable that one should be able to measure with very high precision. 3) There is a prediction from the most fundamental QCD model for the QGP: lattice QCD. 4) Conserved charges are an area where HRG models have explained small-to-large system differences, e.g., strangeness enhancement, as evolving from canonical to grand canonical descriptions, see for example
Refs.~\cite{Vislavicius:2016rwi,Vovchenko:2019kes,Cleymans:2020fsc}. 

However, despite existing measurements of net-charge cumulants~\cite{STAR:2014egu,STAR:2019ans,ALICE:2025mkk} and related observables~\cite{ALICE:2012xnj}, relating experimental measurements to theoretical calculations remains difficult, although there are promising examples~\cite{Alba:2014eba}. Experimental measurements have limited acceptance in both rapidity and transverse momentum, $\pT$. At the same time, electric charge is conserved in a collision and so it therefore does not make sense to measure it over the full $4\pi$ acceptance. Although substantial progress has been made on the phenomenological side in understanding the impact of rapidity coverage, see for example Refs.~\cite{Bzdak:2012ab, Ling:2013ksb, Vovchenko:2020tsr}, it is still not clear how to connect theory and experiment when there is limited transverse momentum coverage~\cite{Vovchenko:2020tsr}.  Recent phenomenological progress has made this connection substantially more quantitative. In particular, Ref.~\cite{Parra:2025fse} developed a framework for the $D$ measure that incorporates the evolution of charge fluctuations, local charge conservation, resonance decays, and experimental acceptance. The aim of the present work is complementary: we ask how the experimental side of this mapping can be exposed in terms of directly measurable, event-averaged observables. We will advocate for using the second-order cumulant as a controlled laboratory for this purpose and show that the charge balance function is critical for understanding the acceptance dependence of net-charge fluctuations, as also stressed in Refs.~\cite{Ling:2013ksb,Pruneau:2019baa}. This is particularly relevant for the interpretation of the $D$ measure~\cite{Asakawa:2000wh,Jeon:2000wg}:
\begin{equation}
\label{eq:D}
    D = 4 \frac{\Kt(\dQ)}{\mNch},
\end{equation}
where \Kt is the second-order charge cumulant defined later, $\dQ = N_{+} - N_{-}$ is the event-by-event charge difference and $N_{\rm ch} = N_{+} + N_{-}$.

\section{Motivation}
\label{sec:motivation}

The ability of nearly ideal relativistic hydrodynamics to describe collective flow in both small and large systems is one of the most important insights into the interactions taking place in the matter created in these collisions. The success of hydrodynamics has also motivated extensive work to understand hydrodynamization~\cite{Strickland:2024moq}, i.e., when hydrodynamics becomes applicable to describe the system. In kinetic theory simulations the matter passes through three stages of equilibration: kinetic (momentum), chemical (quark-to-gluon ratio), and thermal (QGP equation of state), see for example Ref.~\cite{Kurkela:2018xxd}. Hydrodynamics only requires kinetic equilibration and can often even be applied before this~\cite{Romatschke:2017vte}. The PYTHIA shoving model~\cite{Bierlich:2017vhg}, for example, attempts to describe flow by adding interactions between strings without imposing chemical equilibrium. Lund-string shoving has a very distinct signature of negative elliptic flow in ultra-small systems~\cite{Bierlich:2024lmb}. At the same time, it also has a conceptually very distinct signature in large systems: strings remain confined, meaning that charges are not free to move around inside the dense system. Since a meson produced away from the string end points is the result of two $q\bar{q}$ breakings, the quantum numbers and their balancing quantum numbers are locally conserved. Therefore one cannot have long-range balance of quantum numbers carried by light quarks, $Q, S, B$, which are dominantly produced when the system hadronizes. This is not the case for a deconfined QGP where the light quarks carrying the same quantum numbers are, at least in principle, free to roam. It has, for example, been proposed that strangeness enhancement in large systems should in fact be interpreted as the lifting of a canonical suppression in large QGP systems~\cite{Vislavicius:2016rwi,Vovchenko:2019kes,Cleymans:2020fsc}, i.e., that strangeness is canonically suppressed in small systems. 

The possible diffusion of quantum numbers has been investigated via balance functions~\cite{Bass:2000az,ALICE:2013vrb,ALICE:2021hjb,ALICE:2023asw}. Importantly, measured balance functions do not broaden significantly in more central heavy-ion collisions~\cite{ALICE:2013vrb,ALICE:2021hjb}. Instead, they tend to narrow due to radial flow~\cite{Bozek:2004dt}. This does not prove that Angantyr is correct, but one should appreciate that this provides a test that could have falsified both PYTHIA and Angantyr, because significant broadening of balance functions with density is not expected in a Lund string model. It is fundamentally a confined model where charge is locally conserved.

Recent lattice studies have also challenged the simplest picture in which deconfinement immediately implies strong screening of the real part of the static heavy-quark potential~\cite{Bala:2021fkm,Bazavov:2023dci}. In these analyses the real part can remain comparatively vacuum-like while an imaginary part, corresponding to an in-medium width, grows. This does not imply the persistence of Lund-like strings in the QGP, but it illustrates that the microscopic interpretation of deconfinement is more subtle than a simple disappearance of confining interactions.

A useful example of why conserved-charge fluctuations may provide information beyond bulk hydrodynamic observables is a chemically undersaturated plasma, in which the quark abundance is below its equilibrium value~\cite{Shuryak:1992wc,Stocker:2015nka,Basu:2021dzv}. Such a system may be kinetically equilibrated, and therefore admit a hydrodynamic description and a meaningful kinetic temperature, without being in chemical equilibrium. Delayed quark chemical equilibration has also been discussed in connection with thermal-photon phenomenology~\cite{Liu:2012ax,Monnai:2014kqa}. This provides one concrete example in which bulk observables may appear close to thermal while conserved-charge fluctuations need not coincide with equilibrium-QCD susceptibilities. 

\section{The importance of balance functions}
\label{sec:hydro}

\subsection{Susceptibilities and experimental counterparts}

Fluctuations of conserved charges provide direct information on the effective
degrees of freedom and interactions of strongly interacting matter.  In the
grand-canonical ensemble the generalized susceptibilities are derivatives of
the QCD partition function,
\begin{equation}
 \chi^{BQS}_{ijk}
 =
 \frac{1}{VT^3}
 \frac{\partial^{i+j+k}\ln Z}
 {\partial\hat{\mu}_B^i
  \partial\hat{\mu}_Q^j
  \partial\hat{\mu}_S^k},
 \qquad
 \hat{\mu}_X=\mu_X/T .
 \label{eq:susceptibilities}
\end{equation}

For an equilibrated system in an appropriate grand-canonical acceptance,
cumulants of the event-by-event conserved-charge distribution can be related
to the corresponding susceptibilities.  In particular, the experimentally
accessible fourth-to-second cumulant ratio is related to
\begin{equation}
  \frac{\Kf(\dQ)}{\Kt(\dQ)}
  \longleftrightarrow
  \frac{\chiQfour}{\chiQtwo},
  \label{eq:cumulant_mapping}
\end{equation}
where $\dQ = N_{+} - N_{-}$ is the event-by-event charge difference and the cumulants are defined as:
\begin{eqnarray}
	&\Kt(X) =  \langle X^2\rangle - \langle X \rangle^2,\\
	&\Kf(X) =  \langle (X - \langle X \rangle)^4 \rangle - 3\Kt(X)^2.
\end{eqnarray}

\subsection{Second order cumulants and balance functions}

This section starts from the same ideas presented recently in Ref.~\cite{Bierlich:2025pkg}. We define the electric-charge balance function as:
\begin{equation}
\label{eq:balanceXi}
\begin{split}
    B_{Q}(\Delta y) \equiv \frac{1}{N_{\rm trig,+} + N_{\rm trig,-}} \Bigg( & \frac{dN_{+-}}{d\Delta y}(\Delta y) - \frac{dN_{++}}{d\Delta y}(\Delta y) \\
     & + \frac{dN_{-+}}{d\Delta y}(\Delta y) - \frac{dN_{--}}{d\Delta y}(\Delta y) \Bigg),
 \end{split}
\end{equation}
where the charge sign listed as $i$ ($j$) in $dN_{ij}$ denotes the electric charge of the trigger (associated) particle and $N_{\rm trig,+} + N_{\rm trig,-}$ is the total number of triggers. The balance function, $B_{Q}(\Delta y)$, quantifies, on a per-trigger basis, the conditional probability of observing an associated particle with opposite electric charge at $\Delta y$, corrected for combinatorial background by subtracting same-sign pairs. It has a simple meaning because if the total electric charge in a collision were always zero and one had a perfect detector then the integral of the balance function should be one because electric charge is conserved. The charge balance function has been measured by ALICE in \pbpb collisions at the LHC, and it narrows slightly with centrality~\cite{ALICE:2013vrb}, which is consistent with a simple picture in which the underlying microscopic hadronization mechanism remains the same but the narrowing results from increasing radial flow~\cite{Bozek:2004dt}.

The balance function also has a very important role for electric-charge fluctuations. If a perfect detector has a rapidity acceptance much wider than the balance function, then a charged particle produced in the center of that detector will always be balanced by a particle of opposite electric charge ($\dQ = 0$) and therefore never contribute to charge fluctuations. So for a detector with broad rapidity coverage, the measured fluctuations are dominated by particles close to the edges, where close is quantified by the width of the balance function. 

It turns out that it is possible to define \Kt in terms of a balance number, \kQ, which is inspired by the balance function. We define the balance number \kQ:
\begin{equation}
\label{eq:kxi}
\kQ \equiv \frac{1}{{\langle N_{\rm ch} \rangle}}\big( \langle \NXIM\NXIP \rangle - \langle \NXIM(\NXIM - 1) \rangle 
+ \langle \NXIP\NXIM \rangle - \langle \NXIP(\NXIP - 1) \rangle \big),
\end{equation}
and note that this expression is actually much more similar to the standard definition of balance functions~\cite{Bass:2000az}. A value of $\kQ$ close to 1 indicates that almost all electric charges are balanced, while a value near 0 indicates that few are balanced~\footnote{Note that \kQ can be negative for an unphysical system, e.g., if positive and negative particles were never present together in the same event.}. The second-order cumulant can then be rewritten as:
\begin{equation}
\label{eq:k2:approx} 
\begin{split}
  \kappa_{2}(\dQ) & = (1-\kQ)\langle N_{\rm ch} \rangle - \langle \dQ \rangle^2.\\
  & \approx (1 - \kQ)\langle N_{\rm ch} \rangle,
\end{split}
\end{equation}
where the approximation holds when $\langle \dQ \rangle$ is small, which we expect to be true for most measurements at the LHC that do not cover the beam rapidity. Note that Eq.~\ref{eq:k2:approx} is similar to Eq.~35 in Ref.~\cite{Pruneau:2019baa}.

As discussed in a specific context in Ref.~\cite{Bierlich:2025pkg}, \kQ can be estimated with good precision from 
$B_{Q}(\Delta y)$ and $dN_{\rm ch}/dy$ as:
\begin{equation}
\label{eq:kq:approx}
\kQ \approx \frac{\int_{y_{\rm min}}^{y_{\rm max}} dN_{\rm ch}(y')/dy \int_{y_{\rm min}-y'}^{y_{\rm max}-y'} B_{Q}(\Delta y') \, d\Delta y'\, dy'}{\int_{y_{\rm min}}^{y_{\rm max}} dN_{\rm ch}(y')/dy \, dy'},
\end{equation}
where $y_{\rm min}$ and $y_{\rm max}$ are the limits of the detector acceptance. Equation~\ref{eq:kq:approx} assumes that the balance function is approximately independent of the trigger rapidity, $B_Q(\Delta y;y_{\rm trig})\simeq B_Q(\Delta y)$, as one would expect for longitudinal Lund strings. Equation~\ref{eq:kq:approx} makes explicit the single-particle-density and finite-acceptance weighting implicit in the relation between charge fluctuations and balance functions. In the boost-invariant limit, where $dN_{\rm ch}/dy$ is constant and the balance function depends only on $\Delta y$, it reduces to the finite-acceptance form discussed in Ref.~\cite{Ling:2013ksb}. The validity of the approximation used in Eq.~\ref{eq:kq:approx} will be tested for more realistic distributions using PYTHIA.

Since $\langle N_{\rm ch} \rangle$ can of course also be determined from $dN_{\rm ch}/dy$, this means that $\kappa_{2}(\dQ)$ can be determined from two observables with a simple interpretation: the average probability that a charge produced inside the detector acceptance is not balanced by another charge, ($1 - \kQ$), times the total number of charged particles in the acceptance, $\langle N_{\rm ch} \rangle$. This shows that the non-trivial correlation physics is encoded primarily in the balance function, as also pointed out in Refs.~\cite{Ling:2013ksb,Pruneau:2019baa}, and makes it easy to relate to a collision model such as the Lund string model: if we reduce the number of string breakings per unit rapidity we will widen the balance function and \kQ will go down and hence \Kt will go up (assuming $\langle N_{\rm ch} \rangle$ is kept constant). However, it is not trivial to see how this relates to the lattice QCD prediction and this formulation highlights the challenge there. 

Another important point is that $B_{Q}(\Delta y)$ and $dN_{\rm ch}/dy$ are event-averaged observables, meaning that they can be corrected for efficiency and even extrapolated into unmeasured \pT regions. Furthermore, the balance function can be measured over rapidity differences up to twice the detector rapidity coverage. Under the assumption that the balance function is independent of the trigger rapidity, geometrical pair-acceptance effects can be corrected using event mixing. This provides an important experimental advantage for reconstructing the balance function entering Eq.~\ref{eq:kq:approx}.

\section{PYTHIA studies of net-charge fluctuations}
\label{sec:pythia}

\begin{figure}
\centering
    \includegraphics[width=0.5\linewidth]{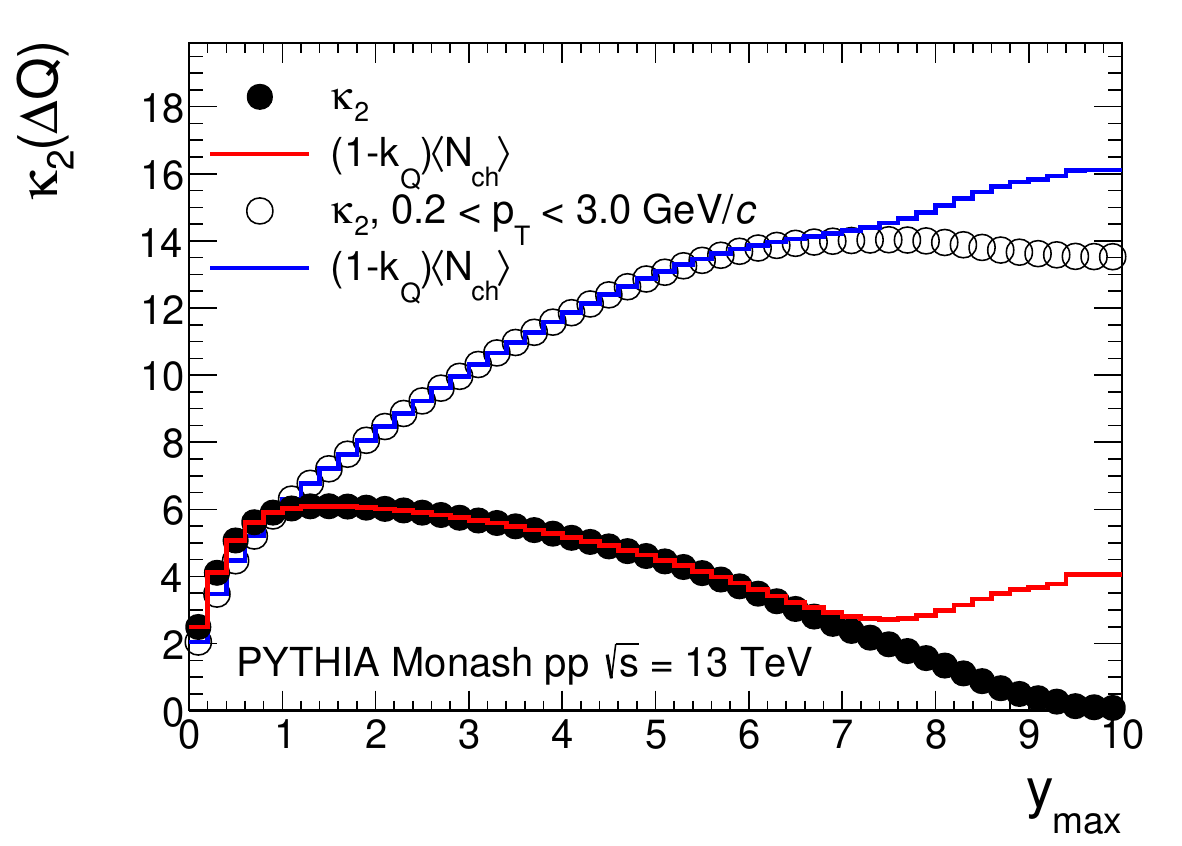}
\caption{\label{fig:k2} PYTHIA \sppt{13} \pp simulations of \Kt as a function of rapidity coverage for a detector with full \pT coverage (solid circles) and one with reduced \pT coverage (open circles). Both are compared with the approximation shown in Eq.~\ref{eq:k2:approx} as lines.}
\end{figure}

In this section the goal is to study the effect of having limited acceptance in rapidity and \pT when measuring $\Kt(\dQ)$. Based on the discussion of the last section it should be clear that we want to have a generator that can generate events in full phase space with a realistic balance function. For this reason we use PYTHIA with the default Monash tune and focus on \pp collisions, which are the fastest to simulate. PYTHIA and HERWIG have previously been used to study higher-order net-particle cumulants in \pp collisions at $\sqrt{s}=13$~TeV as baselines for heavy-ion measurements~\cite{M:2024azy}. Here, rather than using PYTHIA as a baseline prediction for the cumulants themselves, we use its full phase-space information to investigate how the measured cumulants can be reconstructed from experimentally accessible single-particle distributions and balance functions. The acceptance considerations studied here are expected to carry over to heavy-ion collisions. 

The first thing to study is $\Kt(\dQ)$ as a function of the rapidity acceptance. Two configurations will be used. A perfect detector with acceptance $-y_{\rm max} < y < y_{max}$ and one with reduced acceptance in \pT, $0.2 < \pT < \gevc{3.0}$. Figure~\ref{fig:k2} shows the obtained \Kt for both configurations as well as the approximation shown in Eq.~\ref{eq:k2:approx}. The first thing we observe is that the approximation starts to deviate for $y_{\rm max} \gg 6$. This is where the effect of the global charge conservation, $\Delta Q_{global} = 2$, enters and so we observe a deviation because we neglected the second term in Eq.~\ref{eq:k2:approx}. This still leaves a wide acceptance range at the LHC in which global charge conservation is not a complication.

The second thing we observe in Fig.~\ref{fig:k2} is that the reduced \pT acceptance can have an enormous effect for a wide detector. If we think about the approximation, $\Kt(\dQ) \approx (1 - \kQ)\langle N_{ch} \rangle$, then it is clear that the large effect must arise primarily through the change in \kQ. It is also relatively easy to understand why: when the detector does not have full \pT acceptance, the unbalanced charges keep increasing as the $y$ acceptance increases. This is an important point to understand because most measurements are done with limited rapidity acceptance, $y_{max} < 1$,   
where there is only a small difference between the \pT restricted and unrestricted $\Kt(\dQ)$. The small difference in this region can be understood as the reduced efficiency for the \pT restricted measurement reduces both \kQ and $\langle N_{\rm ch} \rangle$ by similar amounts, i.e., the particles we miss for balancing charges will also be lost in the charged-particle measurement. It should be kept in mind that this similarity is only approximate and applies only for limited acceptance.

\begin{figure}
\centering
\begin{minipage}{0.45\linewidth}
    \includegraphics[width=\linewidth]{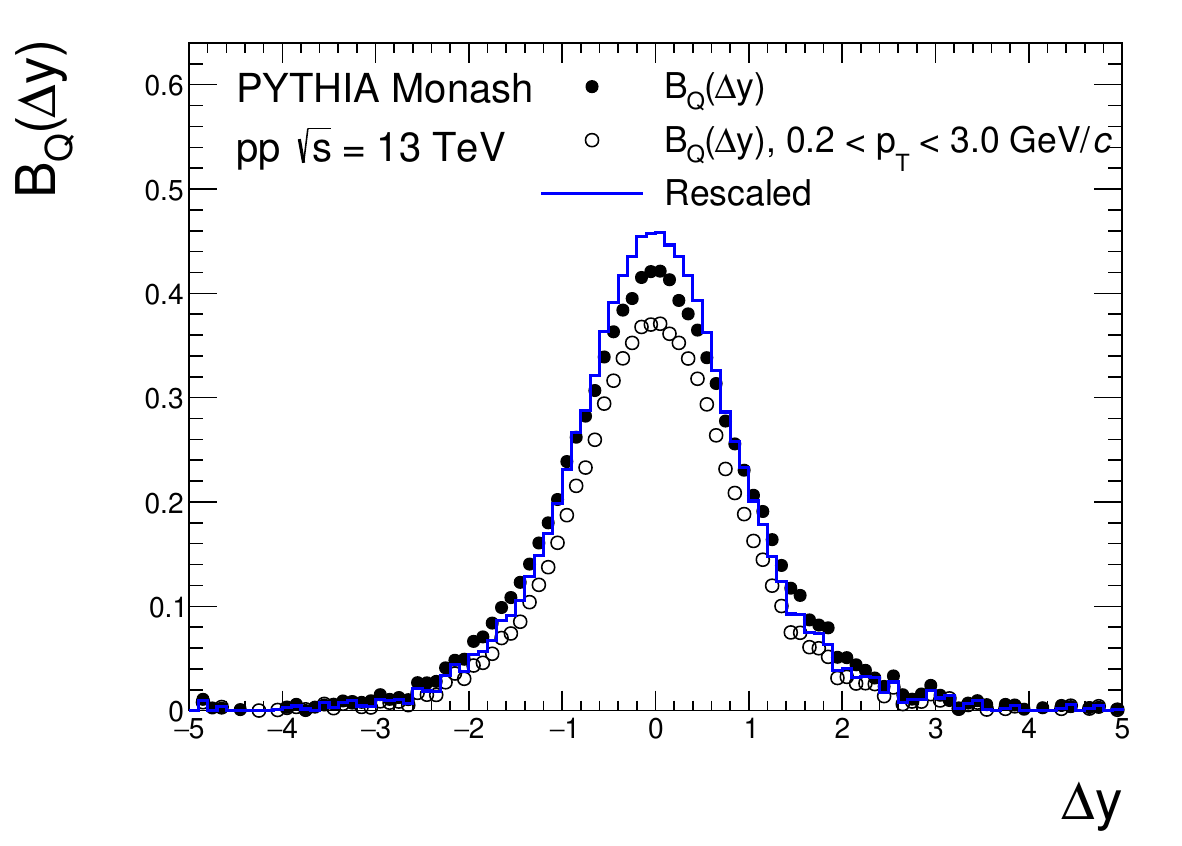}
\end{minipage}
\begin{minipage}{0.45\linewidth}
    \includegraphics[width=\linewidth]{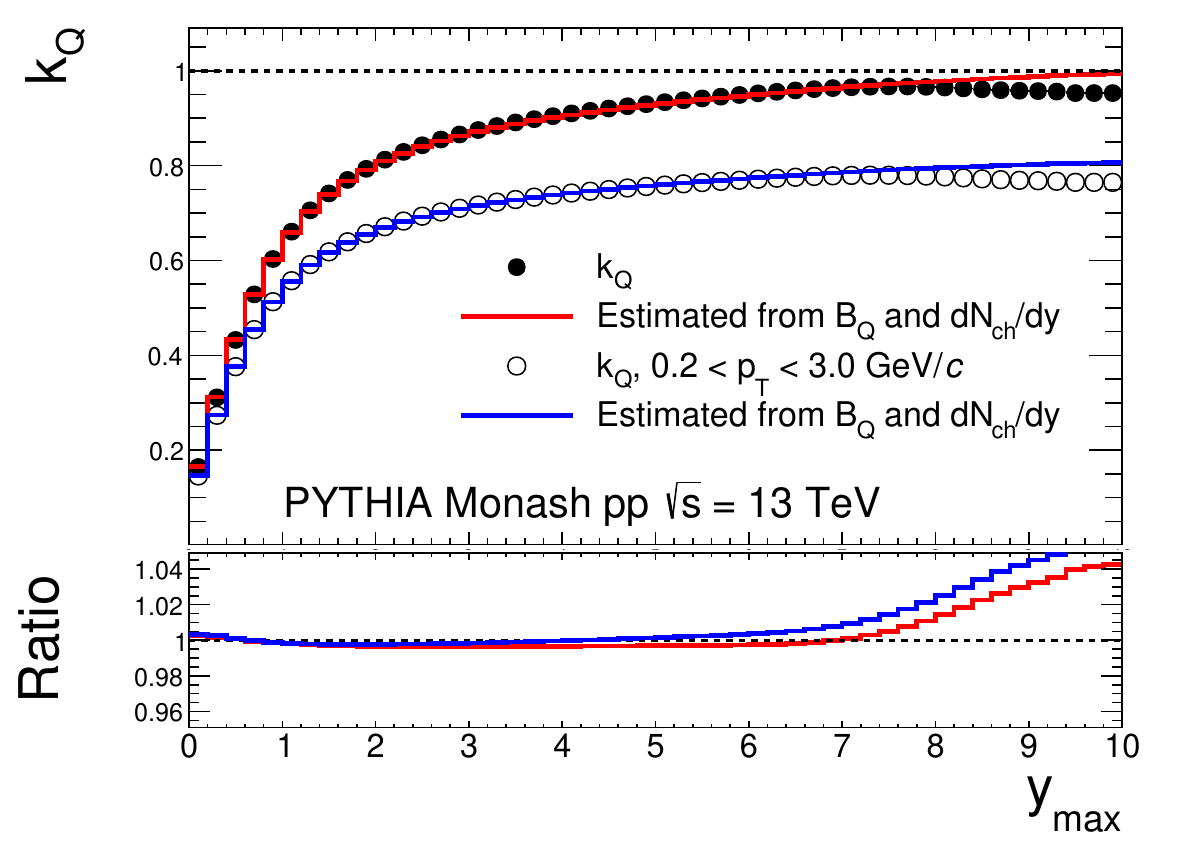}
\end{minipage}
\caption{\label{fig:balance} Left: the balance function for triggers with $|y|<0.2$ for the two detector configurations (solid and open circles) as well as a rescaled version of the one with reduced \pT acceptance to make it easy to compare the shape. Right: \kQ for both configurations as well as an estimate based on Eq.~\ref{eq:kq:approx}. The ratio between the measured and the estimate is shown below.}
\end{figure}

Figure~\ref{fig:balance} left shows the balance function obtained using Eq.~\ref{eq:balanceXi} but accepting only triggers in the rapidity interval $|y|<0.2$. Introducing a \pT restriction reduces the integral of the balance function from ${\approx}1$ to ${\approx}0.8$, but it also changes the shape as can be seen from the rescaled one, where the integral has been fixed to be the same as for the balance function without \pT restriction. We expect that the narrowing of the balance function with the \pT cut is a result of color reconnection, which has a similar effect on \pT spectra as radial flow~\cite{OrtizVelasquez:2013ofg}.

Figure~\ref{fig:balance} right shows \kQ calculated for both configurations. In both cases it first grows quite rapidly before settling into a moderate, asymptotic growth towards the integral of the balance function (1 and 0.8, respectively). As also seen in the figure, Eq.~\ref{eq:kq:approx} does a very good job of reproducing \kQ in regions where one can neglect the global charge. Thus one can derive \kQ from the balance function and $dN/dy$ and it seems that the mid-rapidity balance function is enough to describe the relevant region for these PYTHIA-based studies.   

\begin{figure}
\centering
    \includegraphics[width=0.5\linewidth]{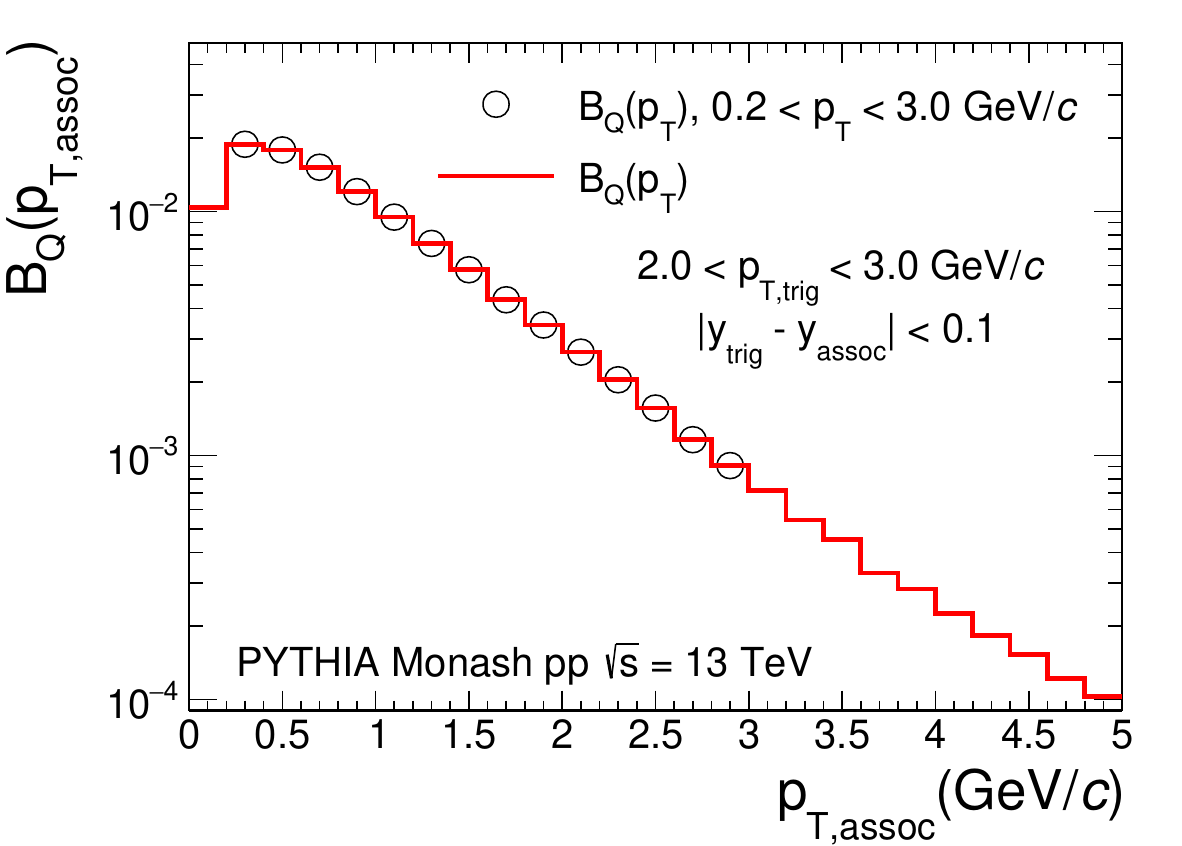}
\caption{\label{fig:balance:pT} The balance function as a function of \pT for $|\Delta y|<0.1$ and triggers with $2 < p_{\rm T,trig} < \gevc{3}$ for the two detector configurations (red line and open circles).}
\end{figure}

So far we have only shown the balance function as a function of the rapidity difference, but it is of course possible to define it instead as a function of associated \pT, which is done next to illustrate how one can extrapolate a measured \pT restricted balance function to the full \pT range. Fig.~\ref{fig:balance:pT} shows the balance function as a function of the associated \pT for fixed triggers with $2 < p_{\rm T,trig} < \gevc{3}$ in the rapidity range, $|\Delta y|<0.1$. As can be seen, this is really a standard \pT spectrum and just like a normal \pT spectrum one can correct it for efficiency and contamination from secondaries and finally extrapolate it to the full ranges, as illustrated with the red curve. This suggests that the extrapolation of balance functions into experimentally inaccessible \pT regions could be approached using techniques similar to those employed for single-particle spectra, although the associated systematic uncertainties would need to be evaluated.

\begin{figure}
\centering
    \includegraphics[width=0.5\linewidth]{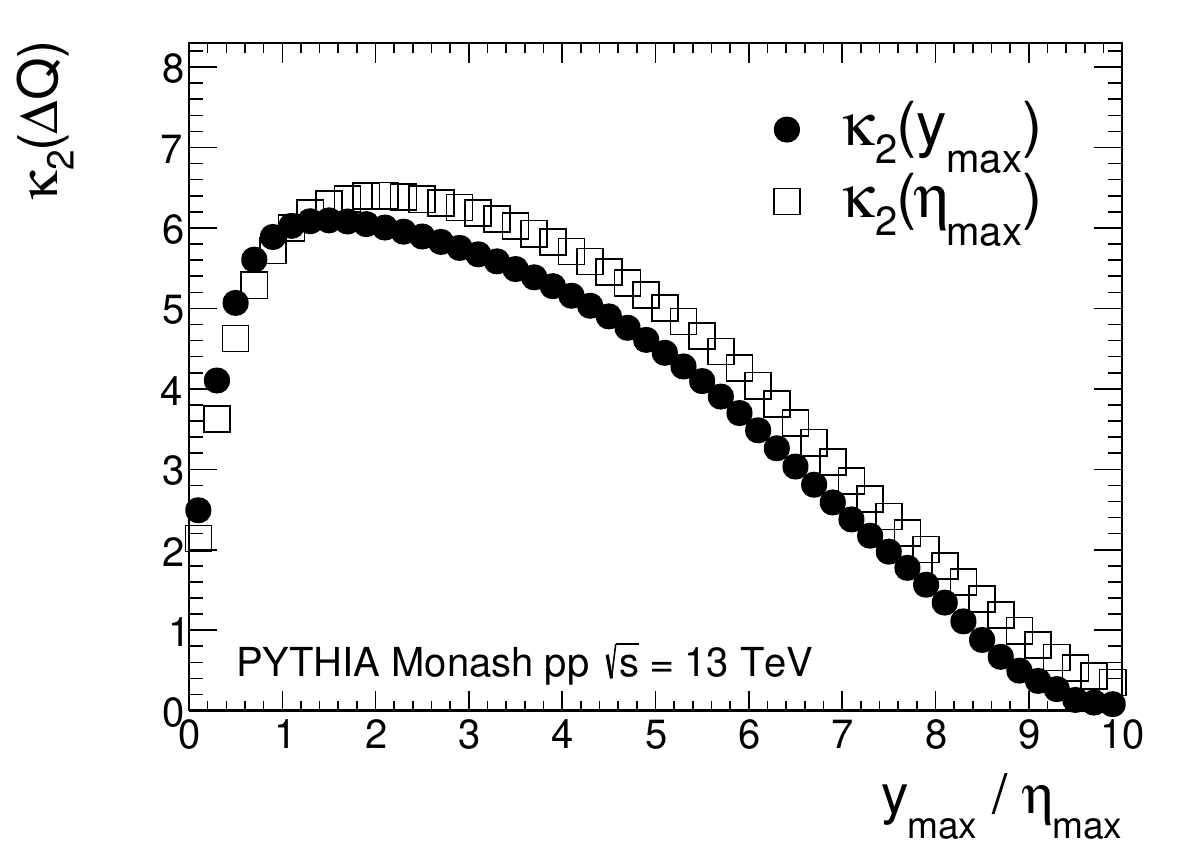}
\caption{\label{fig:k2_y_vs_eta} \Kt as a function of rapidity (solid circles) and pseudorapidity (open squares) coverage for a detector with full \pT coverage.}
\end{figure}

Finally, Fig.~\ref{fig:k2_y_vs_eta} shows a comparison between $\Kt(y)$ and $\Kt(\eta)$. As can be seen, there are significant differences between the two. To be able to do the rapidity measurements one has to do particle identification. This is more tractable for event-averaged quantities and may enable high-precision measurements. Ideally, one should measure both $\Kt(y)$ and $\Kt(\eta)$ because this would also allow comparisons with experiments that do not have particle identification but may have a wider rapidity acceptance.
Given that these measurements are much easier for unidentified particles using pseudorapidity, it would be worthwhile to investigate whether phenomenological comparisons can also be made directly in pseudorapidity, for example using PYTHIA to quantify the mapping between $y$ and $\eta$. I return to this point in the next section.

\section{Discussion}
\label{sec:discussion}

The central result of this study is that the acceptance dependence of \Kt can be exposed in terms of two experimentally accessible, event-averaged observables: $B_Q(\Delta y)$ and $dN_{\rm ch}/dy$. This is complementary to recent phenomenological work on the $D$ measure. In particular, Ref.~\cite{Parra:2025fse} starts from QGP or hadronic charge fluctuations and propagates them through local charge conservation, resonance decays, and experimental acceptance in order to compare with the measured $D$. The present approach starts from the experimental side and asks which measured correlation carries the information that is otherwise compressed into the integrated cumulant.

Using Eq.~\ref{eq:k2:approx}, the $D$ measure can be written exactly as:
\begin{equation}
    D = 4(1-\kQ)-4\frac{\langle \dQ \rangle^2}{\mNch},
\end{equation}
and therefore, for the approximately charge-symmetric midrapidity region at LHC energies:
\begin{equation}
\label{eq:Dkq}
    D \approx 4(1-\kQ).
\end{equation}
Equation~\ref{eq:Dkq} makes the experimental content particularly transparent. The non-trivial correlation information entering $D$ is contained in \kQ, which in turn can be reconstructed from $B_Q(\Delta y)$ and $dN_{\rm ch}/dy$ through Eq.~\ref{eq:kq:approx}. Rather than comparing only the final integrated value of $D$ or \Kt, a comparison of the underlying $B_Q(\Delta y)$ and $dN_{\rm ch}/dy$ can therefore provide substantially more control over acceptance effects and over the microscopic origin of an agreement or disagreement.

This suggests a direct way of matching the present experimental strategy to Ref.~\cite{Parra:2025fse}. The authors' calculation can be used to predict $D$ as a function of acceptance, while measurements of $B_Q$ and $dN_{\rm ch}/dy$ can determine the corresponding \kQ experimentally. It would be particularly useful to investigate whether the correlation functions entering their framework can be projected onto a predicted balance function. Such a prediction could then be confronted directly with measured balance functions, for example Ref.~\cite{ALICE:2013vrb}, instead of only with the acceptance-integrated $D$ measure. We do not assume that $D$ uniquely determines $B_Q$; rather, the aim would be to expose in the phenomenological calculation the same differential correlation information that is directly accessible experimentally.

A second route is to develop event generators that provide a common bridge between equilibrium calculations and detector-level observables. EPOS~4~\cite{Werner:2023jps} provides a full event-generator framework that includes HRG, while Thermal-FIST~\cite{Vovchenko:2019pjl} can both directly evaluate observables within an HRG description and generate event-by-event thermal samples as used in Ref.~\cite{ALICE:2024rnr}. Thermal-FIST is particularly interesting because the same underlying thermal model can be confronted with lattice-QCD quantities and with fluctuation measurements. For the present purpose, however, it would be valuable to extend such thermal sampling to a realistic full event with longitudinal dynamics, charge balancing, resonance decays, and complete phase space. One possibility would be to interface a Thermal-FIST-like thermal source to a full event generator such as EPOS. A framework in which the same underlying model can be evaluated both at the level of equilibrium susceptibilities and after realistic experimental measurements within a constrained acceptance would provide a powerful basis for precision tests of lattice QCD. It could also make controlled comparisons to pseudorapidity-based measurements possible.

The natural next question is whether the same reformulation can be developed for the fourth-order cumulant. For the second order, the relation to a two-particle balance function is especially simple because \Kt is determined by two-particle correlations. In contrast, \Kf contains genuine correlations up to fourth order. Writing $\dQ=N_+-N_-$, one has in terms of joint cumulants
\begin{equation}
\Kf(\dQ)=\Kf(N_+)+\Kf(N_-)-4\kappa_{31}+6\kappa_{22}-4\kappa_{13},
\label{eq:k4joint}
\end{equation}
where $\kappa_{ij}$ denotes the joint cumulant containing $i$ powers of $N_+$ and $j$ powers of $N_-$. Thus a single two-particle balance function cannot in general contain all the information needed for \Kf. The experimental analogue of Eq.~\ref{eq:kq:approx} would require higher-order charge-balance correlations, or an equivalent formulation in terms of multi-particle factorial cumulants. An interesting special case was studied in Ref.~\cite{Altsybeev:2020qnd}, where local production of nearly uncorrelated sources of particle--antiparticle pairs leads to relations in which higher-order net-charge cumulants can be expressed through the second net-charge cumulant and low-order particle-number cumulants. This demonstrates that useful higher-order reformulations are possible under additional assumptions. The more general experimental question is whether one can construct a higher-order analogue of \kQ that retains the transparent acceptance interpretation of Eq.~\ref{eq:kq:approx} without relying on an independent-pair picture. This is potentially important for the lattice-QCD motivation of the present work. The quantity proposed in Ref.~\cite{Borsanyi:2026jbj} is $\chiQfour/\chiQtwo$, whose experimental counterpart involves $\Kf/\Kt$. Establishing a differential, acceptance-aware representation of \Kf analogous to the balance-function representation of \Kt would therefore provide a direct route toward controlling the experimental corrections needed for a precision comparison with lattice QCD. The second-order study presented here can be viewed as a first step toward such a construction.

\section{Conclusions}

Recent lattice-QCD calculations of electric-charge susceptibilities, together with phenomenological developments that propagate charge fluctuations to finite experimental acceptance, motivate a renewed precision program for net-charge fluctuations at the LHC. In this work we have approached the problem from the experimental side. For the second-order cumulant, the acceptance dependence can be expressed through the charge balance function and the charged-particle rapidity density. A PYTHIA study demonstrates that this representation reproduces the directly calculated $\Kt(\dQ)$ over the experimentally relevant rapidity range and makes the effects of restricted $\pT$ acceptance explicit.

The main advantage of this formulation is that $B_Q(\Delta y)$ and $dN_{\rm ch}/dy$ are event-averaged, differential observables. They therefore expose information that is compressed into the integrated $D$ measure and offer additional handles for efficiency corrections, extrapolations, and comparisons with phenomenological calculations such as Ref.~\cite{Parra:2025fse}. Extending this strategy to the fourth-order cumulant is less direct because genuine higher-order correlations enter, but existing work on local charge conservation~\cite{Altsybeev:2020qnd} shows that useful relations can emerge under additional assumptions. Developing a general, experimentally transparent analogue of the balance-function construction for $\Kf/\Kt$ would be an important step toward a precision comparison of measured net-charge fluctuations with the new lattice-QCD prediction for $\chiQfour/\chiQtwo$.

\section*{Acknowledgment}

The author would like to thank the participants in the ``QCD challenges from pp to AA collisions'' International Workshop in Torino in July 2026 for inspiring presentations and discussions. The author also thanks Alice Ohlson for carefully reading the manuscript and for useful comments and suggestions.\\

Support from the following research grant is gratefully acknowledged: Vetenskapsrådet contract 2021-05179.

\bibliographystyle{apsrev4-1}
\bibliography{bibliography} 

\end{document}